# The importance of hole concentration in establishing carrier-mediated ferromagnetism in Mn doped Ge

Shengqiang Zhou[1,2,a], Danilo Bürger[1], Wolfgang Skorupa[1], Peter Oesterlin[3], Manfred Helm[1], and Heidemarie Schmidt[1]

[1] Institute of Ion Beam Physics and Materials Research, Forschungszentrum Rossendorf, P.O. Box 510119, 01314 Dresden, Germany
[2] State Key Laboratory of Nuclear Physics and Technology, School of Physics, Peking University, Beijing 100871, China
[3] INNOVAVENT GmbH, Bertha-von-Suttner-Str. 5, 37085 Göttingen, Germany

Electronic mail: s.zhou@fzd.de.

**Abstract:**
In the present work, we have prepared Mn-doped Ge using different annealing approaches after Mn ion implantation, and obtained samples with hole concentrations ranging from $10^{18}$ to $2.1 \times 10^{20}$ cm$^{-3}$, the latter being the highest reported so far. Based on the magnetotransport properties of Mn doped Ge, we argue that the hole concentration is a decisive parameter in establishing carrier-mediated ferromagnetism in magnetic Ge.

Mn doped GaAs (Ref. [1]) and ZnTe (Ref. [2]) are considered as the prototype diluted ferromagnetic semiconductors (FMS). In both systems, a large enough hole concentration is a prerequisite to establish the carrier-mediated ferromagnetism, which allows the electrical control over magnetism.[3] Practically, it is highly desirable to realize a FMS compatible with silicon technology. Whereas attempts to fabricate magnetically doped silicon (Si:Mn) have been rather discouraging,[4] germanium appears to be a promising candidate due to the well-accepted substitutional occupation of Mn in the Ge matrix.[5,6,7,8] The substitutional Mn ions produce double-acceptor levels and supply holes.[9]

The ferromagnetism in Ge:Mn has been nonquantitatively but plausibly explained by the formation of bound magnetic polarons (BMP).[10,11,12,13] According to the model by Kaminski and Das Sarma,[14] the percolation of BMP over the entire sample depends on the temperature and on the hole concentration. If GaAs:Mn can be any guide, a large enough hole concentration is required to establish carrier-mediated ferromagnetism.[15,16] For GaAs:Mn with 8.5% Mn, the critical hole concentration at 10 K is estimated to be $3 \times 10^{19}$ cm$^{-3}$ (Ref. [15]). In Fig. [1] we summarize the hole concentration realized until now in Ge:Mn.[10,17,18,19,20,21,22,24] One can see that most of the values are well below the threshold value of $3 \times 10^{19}$ cm$^{-3}$. Basically, this survey explains the lack of a correlation between magnetization and magnetotransport in Ge:Mn,[10,17,18,19,24,25] which is a hallmark of FMS.[2,26] Zhou et al.[27] shows that the positive magnetoresistance (MR) and the anomalous Hall resistance in Ge:Mn are likely related with a two-bandlike conduction. Furthermore, a correlation between magnetization and magnetotransport in Ge:Mn was observed at low temperatures when the hole concentration is large enough.[28]

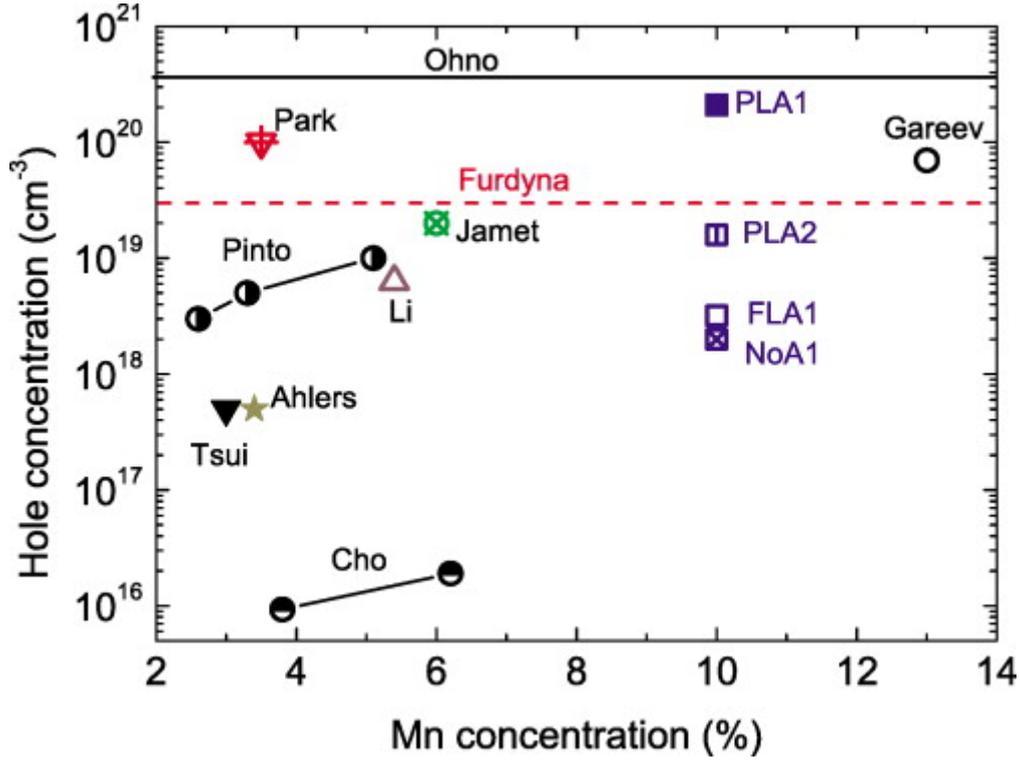

Fig 1. An overview of literature data showing the hole concentration vs Mn concentration in Ge:Mn. The four samples with 10% Mn are from this work, in which the hole concentration is determined from the ordinary Hall effect measured at 5 K. The other samples are from Park (Ref. [10]), Jamet (Ref. [17]), Tsui (Ref. [18]), Gareev (Ref. [19]), Pinto (Ref. [20]), Li (Ref. [12]), Cho (Ref. [21]), and Ahlers (Ref. [22]), for which the hole concentration is reported at room temperature. Except for the work by Cho (Ref. [21]), all other Ge:Mn samples have been prepared by low-temperature molecular-beam epitaxy. The solid black line labeled "Ohno" is the hole concentration in a GaAs:Mn FMS with $T_C$ K = 105 (Ref. [23]). The dashed red line "Furdyna" is the lowest hole concentration leading to ferromagnetism for GaAs:Mn (Ref. [15]).

In this article we present the magnetic and magnetotransport properties of Mn-doped Ge with hole concentrations ranging from $10^{18}$ to over $10^{20}$ cm$^{-3}$, which is achieved by applying different annealing approaches after ion implantation. We argue that the hole concentration is a decisive parameter in establishing the carrier-mediated ferromagnetism in Mn-doped Ge.

Nearly intrinsic, n-type Ge(001) wafers were implanted with Mn ions. The implantation energy and fluence were 100 keV and 30 keV, and $5\times10^{16}$ cm$^{-2}$ and $1\times10^{16}$ cm$^{-2}$, respectively, resulting in a boxlike distribution of Mn ions with concentration around 10% over a depth of 100 nm. During implantation the wafers were flow-cooled with liquid nitrogen to avoid the formation of any Mn-rich secondary phase. A control wafer was implanted at 300 °C (100 keV, $5\times10^{16}$ cm$^{-2}$), resulting in the formation of Mn$_5$Ge$_3$ crystalline precipitates during implantation.[29]

Flash lamp annealing (FLA) was performed with a pulse duration of 3 ms. The energy density was from 53.6 to J/cm 65.4$^2$, which is enough for the implanted layer regrowth.[30] Pulsed laser annealing (PLA) was performed at Innovavent GmbH using a laser ASAMA 80–8 and an optical system VOLCANO. The pulse duration was 300 ns at a wavelength of 515 nm. The sample surface was scanned by a stripelike laser beam mm×40 (2$\mu$m) with a frequency of 50 kHz. The introduced energy density amounted to 1.0 and J/cm 1.5$^2$, which is enough for the liquid phase epitaxial regrowth. PLA has been used to fabricate GaAs:Mn and GaP:Mn FMS.[31,32,33] Magnetic properties were measured with a superconducting quantum interference device Quantum Design Magnetic Property Measurement System (MPMS) magnetometer. Magnetotransport properties were measured with a magnetic field applied perpendicularly to the film plane in van der Pauw geometry.

All samples presented in this work are listed in Table 1. Figure 2a shows the zero-field cooling (ZFC)/field cooling (FC) magnetization curves. In general, all samples show an irreversibility in magnetization depending on their thermal history. This is the typical behavior of magnetic clusters. The low-concentration samples NoA1 and FLA1 reveal similar ZFC/FC curves with a peak at around 270 K, which corresponds to $Mn_5Ge_3$ precipitates. $Mn_5Ge_3$ has been confirmed by x-ray diffraction (not shown) in sample NoA1, that has been implanted with Mn at 300°C (Table 1). FLA with a pulse duration of 3 ms, as employed for sample FLA1, can effectively prevent the formation of MnAs precipitates in GaAs.[34] In sharp contrast, samples PLA1 and PLA2 annealed by a laser pulse have a large hole-concentration and display ZFC/FC curves significantly different from samples NoA1 and FLA1. The most noticeable feature is the ZFC peak in the temperature range from 60 to 70 K, which is related to Mn-rich regions in Ge.[10,11,12] The inset of Fig. 2a shows the remanent magnetization versus temperature for samples FLA1 and PLA1. Despite the large difference in ZFC/FC curves, a magnetic phase is observed for both samples at low temperature. This ferromagnetic phase is related with the coupling between diluted Mn ions in Ge.[11,13,17,35,36]

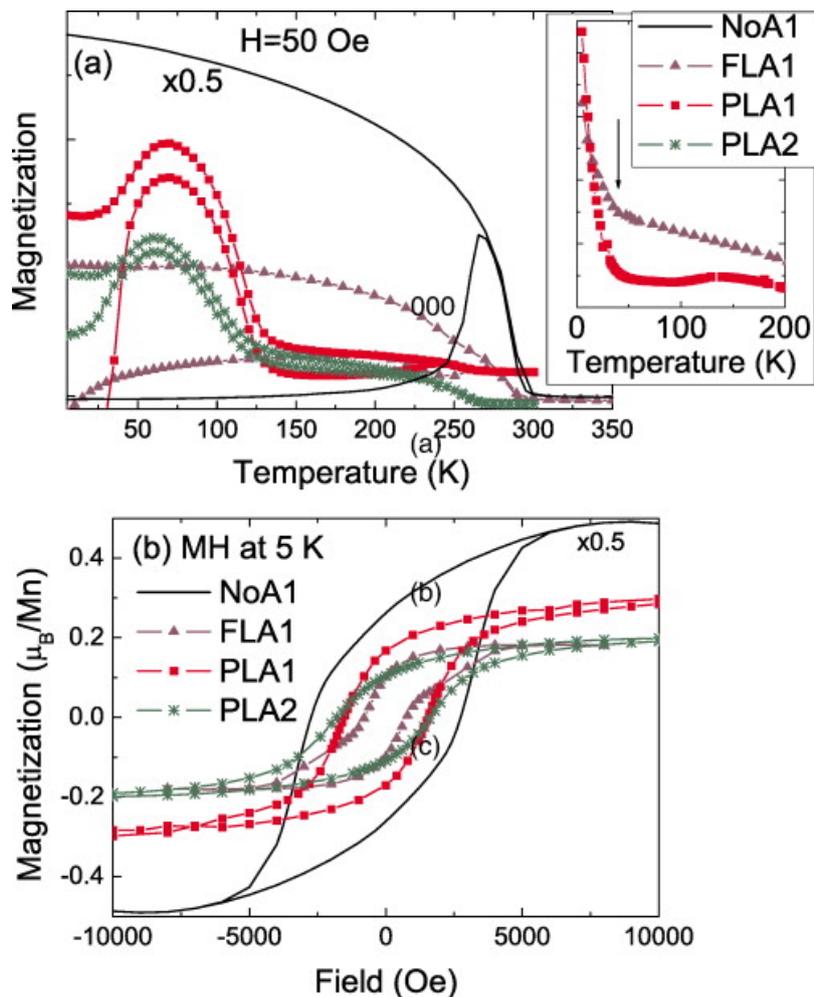

Fig 2. Magnetic properties of Mn-implanted Ge after different thermal processing: (a) Temperature-dependent magnetization after ZFC (lower branches) and after field cooling FC (upper branches). The inset shows the remanent magnetization vs temperature, and a low temperature magnetic phase is indicated by the arrow. The negative ZFC magnetization of sample PLA1 is due to the residual field in the superconducting magnet employed during magnetization measurements. (b) Magnetization at 5 K. Sample NoA1 containing $Mn_5Ge_3$ clusters exhibits the largest moment, while the other samples do not differ strongly.

Note that the ZFC/FC curves of samples PLA1 and PLA2 also reveal a weak ferromagnetic component up to 260 K. This may be due to some small residual $Mn_5Ge_3$ precipitates. We have performed synchrotron x-ray diffraction measurements on both samples (not shown). PLA1 reveals a small peak close to $Mn_5Ge_3$(002) but PLA2 reveals no extra peak besides the Ge phase. $Mn_{11}Ge_8$ has been observed in none of the samples.

Figure 2b shows the field-dependent magnetization at 5 K. Sample NoA1 exhibits the largest saturation magnetic moment, while the magnetization of sample PLA2 is smaller than that of sample PLA1 and comparable to sample FLA1.

As shown in Fig. 2, samples PLA1 and PLA2 reveal similar magnetic properties. Moreover, the low temperature (below 30 K) ferromagnetic phase has been clearly observed in all of the investigated samples by us [see the inset of Fig. 2a], as well as by other groups.[11,13,17,35,36] However, we observed that the magnetotransport at low temperatures is much different. Figure 3a shows the MR at 5 K. Samples PLA1 and PLA2 have a negative MR but at large field the positive MR overcomes the negative one for sample PLA2. Figure 3b shows the Hall effect of different samples. The ordinary and anomalous Hall terms depend linearly on field and magnetization, respectively. At high fields, all samples exhibit a linear behavior, which allows the determination of the hole-concentration by a linear fitting of the Hall curve (see Table 1 and Fig. 1). The zoom-in of the MR and Hall curves in the low field region is shown in Figs. 3c,3d, respectively. Only samples PLA1 and PLA2 exhibit clear hysteretic MR and AHE curves (weak for sample PLA2). Compared with literature data (Fig. 1), the sample PLA1 reaches the largest hole-concentration. If the same set of carriers participate in ferromagnetic coupling and transport, one expects a large AHE, corresponding to the magnetization of the sample.[2,26] The hole concentration is the critical parameter for carrier-mediated ferromagnetism throughout the Ge:Mn sample. Only a large enough hole concentration (PLA1) gives rise to the carrier-mediated ferromagnetism throughout the Ge:Mn sample. The hole concentration of the other samples (PLA2, FLA1, and NoA1) is too small to mediate the ferromagnetism in Ge:Mn. A theoretical picture on the antiferromagnetic coupling between holes and Mn ions acting as double acceptors is given in Ref. 28.

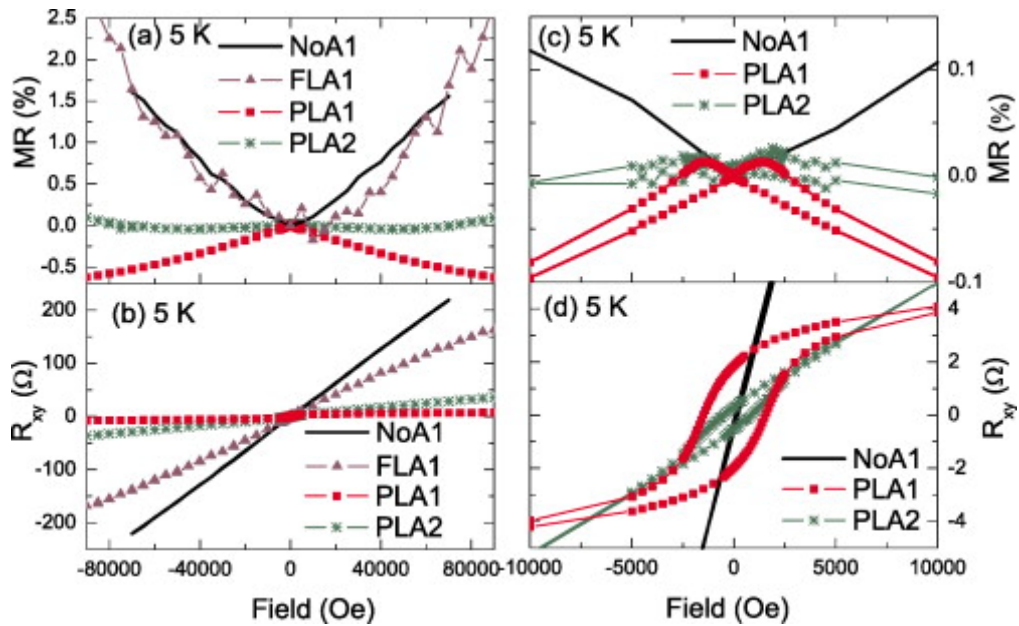

Fig 3. Magnetotransport properties of Mn-implanted Ge at 5 K: (a) MR, Samples PLA1 and PLA2 show a negative MR. (b) Hall resistance ($R_{xy}$). [(c) and (d)] Zoom-in of the low-field region of the MR and Hall curves. Samples PLA1 and PLA2 show hysteresis in MR and Hall curves, while samples NoA1 and FLA1 do not.

In a summary, we present the magnetic and magnetotransport properties of a group of Ge:Mn samples with the hole concentration ranging from $10^{18}$ to $10^{20}$ cm$^{-3}$. The hole concentration is the critical parameter to establish carrier mediated ferromagnetism in Ge:Mn, as seen for GaAs:Mn.[1,15,16] A high-concentration codoping with a shallow acceptor[30] may allow to increase the hole-concentration further, possibly resulting in a dramatically increased Curie temperature.

## ACKNOWLEDGMENTS

Financial support from the Bundesministerium für Bildung und Forschung (Grant No. FKZ13N10144) is gratefully acknowledged. The authors acknowledge the support of Thomas Schumann for FLA.

Table I. Sample identification, Mn ion implantation temperature, annealing conditions, and hole-concentration. NoA1: implanted at a high temperature and nonannealed; FLA1: flash lamp annealed; PLA1 and PLA2: pulsed laser annealed.

| Sample | Implantation temperature (°C) | Annealing conditions | Hole concentration (cm$^{-3}$) |
|---|---|---|---|
| NoA1 | 300 | ... | $2.0 \times 10^{18}$ |
| FLA1 | −40 | 3 ms, Jcm$^{-2}$ 59.4 | $3.2 \times 10^{18}$ |
| PLA2 | −40 | 300 ns, Jcm$^{-2}$ 1.0 | $1.6 \times 10^{19}$ |
| PLA1 | −40 | 300 ns, Jcm$^{-2}$ 1.5 | $2.1 \times 10^{20}$ |